\documentclass{article} 
\usepackage{color,amsmath,graphicx,amssymb,url,calc}

\usepackage{fancyhdr}

\begin{document}

\textbf{\Large On beta-skeleton automata with memory}\par
\vspace{0.6cm}
\centerline{Ram\'on Alonso-Sanz$^{(1)}$ and Andrew Adamatzky$^{(2)}$}
\par\vspace{0.5cm}\noindent
{$^1$~ETSI Agronomos (Estadistica, GSC), C. Universitaria 28040, Madrid, Spain\par
\centerline{{\tt ramon.alonso@upm.es}}\par
$^2$~University of the West of England, Bristol BS16 1QY, United Kingdom \\
\centerline{{\tt andrew.adamatzky@uwe.ac.uk}}
\date{}

\begin{abstract}
A $\beta$-skeleton is a  proximity undirected graph whose connectivity is
determined by the  parameter $\beta$\,. We study $\beta$-skeleton automata where every node is a finite state
 machine taking two states, and updating its states depending on the states of  adjacent automata-nodes. We allow
 automata-nodes to remember their previous states. In computational experiments we study  
how memory affects the global space-time dynamics on $\beta$-skeleton automata.
\end{abstract}


\pagestyle{fancy}
\lhead{\footnotesize Alonso-Sanz~R. and Adamatzky~A. On beta-skeleton automata with memory.\\ Journal of Computational Science 2 (2011) 1, 57--66. }
\chead{}
\rhead{}

\section{Introduction}

In computational geometry and geometric graph theory, a $\beta$-skeleton or beta skeleton is an undirected proximity  graph
defined from a set of points in the Euclidean plane. Two points $p$ and $q$ are connected by an edge whenever their $\beta$ neighborhood  is empty~\cite{Kirk}\,. In lune-based $\beta$-skeletons, the $\beta$-neighborhood is defined as\,:
 the intersection of two circles of radius $d(p,q)$/2$\beta$ that pass through $p$ and $q$, if $\beta \in [0,1]$\,;
the intersection of two circles of radius $\beta d(p,q) /2$ centered at the points
$(1-\beta/2)p+(\beta/2)q$ and $(\beta/2)p+(1-\beta/2)q$,  if $\beta \ge 1$\,.
\par
Figure~\ref{fig:exbetas} shows three $\beta$-skeletons with increasing $\beta$ parameter value, based on the same ten nodes.
The figure shows also the $\beta$-neighborhoods of the node labeled 1 (upper-right). In the transition from $\beta$=0.9 to $\beta$=1.0,
the 1-node loses its 1-6 and 1-8 links, whereas  from $\beta$=1.0 to $\beta$=1.5 it loses the 1-5 link.


\begin{figure}[!htb]\centering
\includegraphics[width=1.0\textwidth,draft=false]{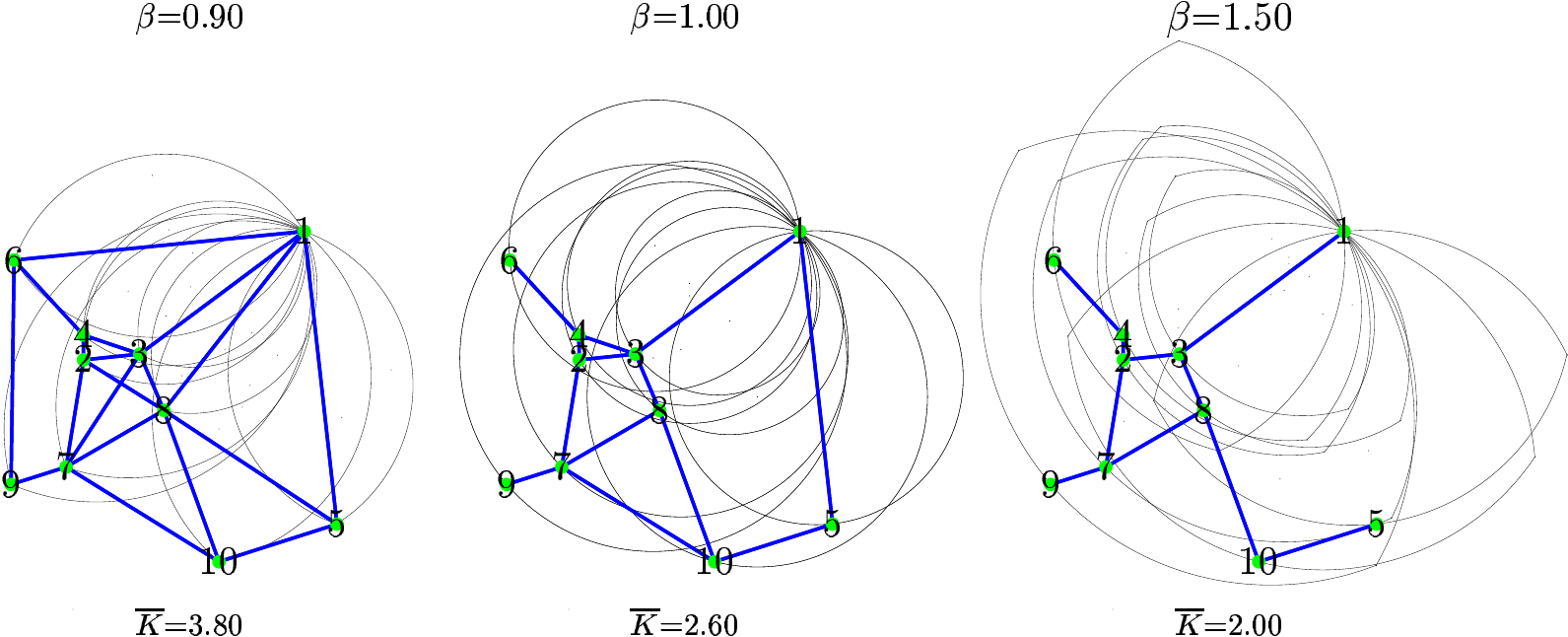}
\caption{Three $\beta$-skeletons on the same ten nodes.}\label{fig:exbetas}
\end{figure}

$\beta$-Skeletons belongs to a  family of proximity graphs. which are  monotonously parameterised by parameter the $\beta$.
The structure of proximity graphs represents a wide range of natural systems and is applied in  many fields of science.
Few examples include geographical variational analysis~\cite{gabriel_1969,matula_1980,sokal_2008},
evolutionary biology~\cite{magwene_2008}, simulation of epidemics~\cite{toroczkai_2008}, study of
percolation~\cite{billiot_2010} and magnetic field~\cite{sridharan_2010}, design of ad hoc wireless networks~\cite{li_2004,song_2004,santi_2005,muhammad_2007,wan_2007}. Thus developing and analysing computational models
of spatially-extended systems on proximity graphs will shed a light onto basic mechanisms of activity propagation on
natural systems.

Automata on $\beta$-skeletons were originally introduced in \cite{Ada} and studied in a context of excitation dynamics.
In present paper we develop ideas of \cite{Ada} along lines of memory-enriched automata and global dynamics.
The paper is structured as follows. In Sect.~\ref{automata} we define automata on $\beta$-skeletons.
Global dynamics on automata for $\beta \geq 1$ (planar graphs) are studied in Sects.~\ref{beta1} and \ref{betamore1}\,,
whereas the $\beta <1$ (non-planar graphs) is studied in Sect.~\ref{betaless1}\,. Section~\ref{weightedmemory} demonstrates the effects
of weighted memory on the behaviour of automata networks. Automata networks with non-parity rules are briefly
tackled in Sect.~\ref{otherrules}. Possible applications of our computational findings are discussed
in Sect.~\ref{applications}.

\section{Automata on beta-skeletons}
\label{automata}

In the automata on beta-skeletons studied here, each node is characterized by an internal state whose value
belongs to a finite set. The updating of these states is made simultaneously ({\it \`a la} cellular automata)
according to a common local transition rule involving only the neighborhood of
each node \cite{Ada}. Thus, if $\sigma ^{(T)}_{i}$ is taken to denote the state value of
node \textit{i} at time step $T$, the site values evolve by iteration of the
mapping : $\sigma ^{(T+1)}_{i} = \phi\Big( \{\sigma^{(T)}_{j}\}\in \mathcal{N}_{i} \Big)$ ,
where $~\mathcal{N}_{i}$ is the set of nodes in the neighborhood of \textit{i} and $\phi$ is an arbitrary function which specifies
the  automaton $rule$. This article deals with two possible state values at each site: $\sigma \in \{0,1\}$,
and the  parity rule\,:  $\sigma^{(T+1)}_{i}=\displaystyle\sum_{j\in\mathcal{N}_{i}}\sigma^{(T)}_{j} \mod 2$\,.
 Despite its formal simplicity, the parity rule may exhibit complex behaviour \cite{Julian}.
\par
In the Markovian approach just outlined (referred as {\it ahistoric}),
 the transition function depends on the neighborhood configuration of the nodes
only at the preceding time step. Explicit historic memory can be embedded in the dynamics by featuring every node by a mapping of
 its states in the previous time steps. Thus, what is
here proposed is to maintain the transition function $\phi$ unaltered, but make it act on the nodes featured by a trait state
obtained as a function of their previous states: $\sigma^{(T+1)}_{i}= \phi \Big(\{s^{(T)}_{j}\}\in \mathcal{N}_{j}\Big)$,
$s^{(T)}_{j}$ being a state function of the series of states of the node $j$ up to time-step $T$. We will consider here
the most frequent state (or {\it majority}) memory implementation. Thus, with unlimited trailing memory\,:
$s^{(T)}_{i}\!=mode\,\big(\sigma^{(T)}_i,\sigma^{(T-1)}_i,\ldots,\sigma^{(\mathrm{1})}_{i}\big)$, whereas
with memory of the last $\tau$ state values\,:
$s^{(T)}_i= mode\big( \sigma^{(T)}_i, \sigma^{(T-1)}_i, \ldots, \sigma^{(\top)}_i\big)\,$, with
$\top=\max (1,T-\tau+1)$\,. In the case of equality in the number of time-steps that a node was 0 and 1, the last state is kept, in which
case memory does not really actuate. This lack of effect of memory explains the lower effectiveness of even size $\tau$-memories
found in the results presented below.
\par

A FORTRAN code working in double precision has been implemented to perform computations. The core of the code is given 
in Table~\ref{tab:core}. The wiring of the $n$ nodes is encapsulated in a $n \times n$ adjacency matrix named ADJ, so that
the {\tt iadd} variable adds the states of the nodes linked to the generic node $i$. Thus, the parity rule is implemented by means
of the {\tt mod 2} operation on the {\tt iadd} variable,  generating {\tt NEW(i)}, i.e., $\sigma_i^{(T+1)}$\,.
Previously, the {\tt MEMORY} subrutine implements the {\tt itau}-majority memory computing. {\tt MEMORY} provides
 the {\tt MODE(i)} trait states, i.e., $s_i^{(T)}$\,, on which the  {\tt iadd} is calculated.

\begin{table}[!htb]
{\tt \footnotesize
\par\hspace{1cm}
         DO it=1,maxit\par\hspace{1.5cm}
           call MEMORY(MODE,NEW,ISIG,n,it,maxit,itau)\par\hspace{1.5cm}
           DO i=1,n;iadd=0\par\hspace{2cm}
              do j=1,n;if(ADJ(i,j)==0)cycle\par\hspace{3cm}
               iadd=iadd+MODE(j)\par\hspace{2cm}
              enddo\par\hspace{2cm}
                NEW(i)=mod(iadd,2)\par\hspace{1.5cm}
            ENDDO\par\hspace{1cm}
         ENDDO \par
\par\vspace{0.2cm}\hspace{1cm}
      subroutine MEMORY(MODE,NEW,ISIG,n,it,maxit,itau)\par\hspace{1.5cm}
        integer MODE(n),NEW(n),ISIG(n,maxit)\par\hspace{1.5cm}
         ISIG(:,it)=NEW;MODE=NEW\par\hspace{1.5cm}
          itin=max(1,it-itau+1);itaux=min(it,itau)\par\hspace{1.5cm}
          do i=1,n; inn=sum(ISIG(i,itin:it))\par\hspace{2.5cm}
               if(2*inn>itaux)MODE(i)=1\par\hspace{2.5cm}
               if(2*inn<itaux)MODE(i)=0\par\hspace{1.5cm}
          enddo\par\hspace{1cm}
      END
}
\caption{The core of the FORTRAN progam.}\label{tab:core}
\end{table}

\subsection{The $\beta$=1 case.}\label{beta1}

Figure~\ref{fig:exb1} shows the initial evolving patterns of a simulation  of the parity rule on a  $\beta$=1 skeleton
(or Gabriel maps) with $N=10^2$ nodes distributed at random in a unit square. Red squares denote node state values equal one,
black squares denote zero state values.  The effect of endowing nodes with memory of the $majority$ of the last three states is shown  at $T$=4\,. The encircled node exemplifies the initial effect of memory. Dynamics of perturbation spreading looks unpredictable and quasi-chaotic due to existence of long-range connections.

\begin{figure}[!htb]\centering
\includegraphics[width=1.0\textwidth,draft=false]{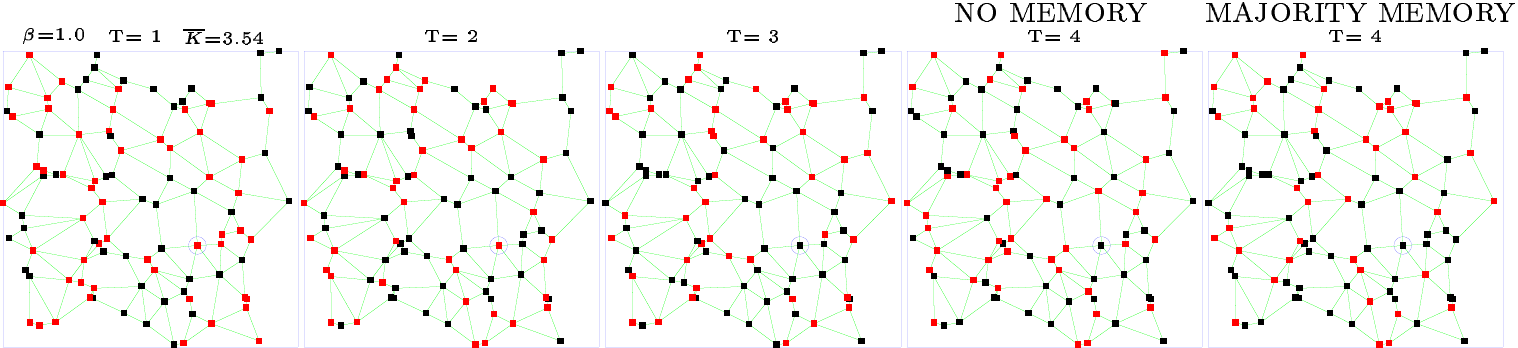}
\caption{A simulation up to $T$=4 of the parity rule on  a $\beta$=1 skeleton,  $N=100$.
Red squares denote node state value equal one, black squares denote zero state values. }\label{fig:exb1}
\end{figure}

\par
Figure~\ref{fig:crb1} shows the evolution of the changing rate (the Hamming distance between two consecutive patterns)
in eleven different $\beta$=1 simulations based in one thousand nodes distributed at random in a unit square.
The red curves correspond to the ahistoric simulations, in which case the parity rule
exhibits a very high level of changing rate, oscillating around 0.5\,. Figure~\ref{fig:crb1} shows also the effect on
the changing rate of endowing nodes with memory of the last $\tau$ state values (blue lines). The inertial effect of memory  tends to reduce the changing rate compared to the ahistoric model, particularly when $\beta$ is odd. With high memory charges,
such as $\tau$=19 in the lower left panel, the changing rate tends to vary in the long term in the $[0.1,0.2]$ interval,
after an initial {\it almost-oscillatory} behaviour which ceases by $T>\tau$+1=20.
With unlimited trailing memory (lower right panel), this {\it oscillatory} pattern is never truncated, so that a rather unexpected
quasi-oscillatory behaviour turns out with full memory. With no exception, the proportion of node states
having  one given state value ({\it density}), oscillates near to 0.5 regardless of the model considered.

\begin{figure}[!htb]\centering
\includegraphics[width=1.0\textwidth,draft=false]{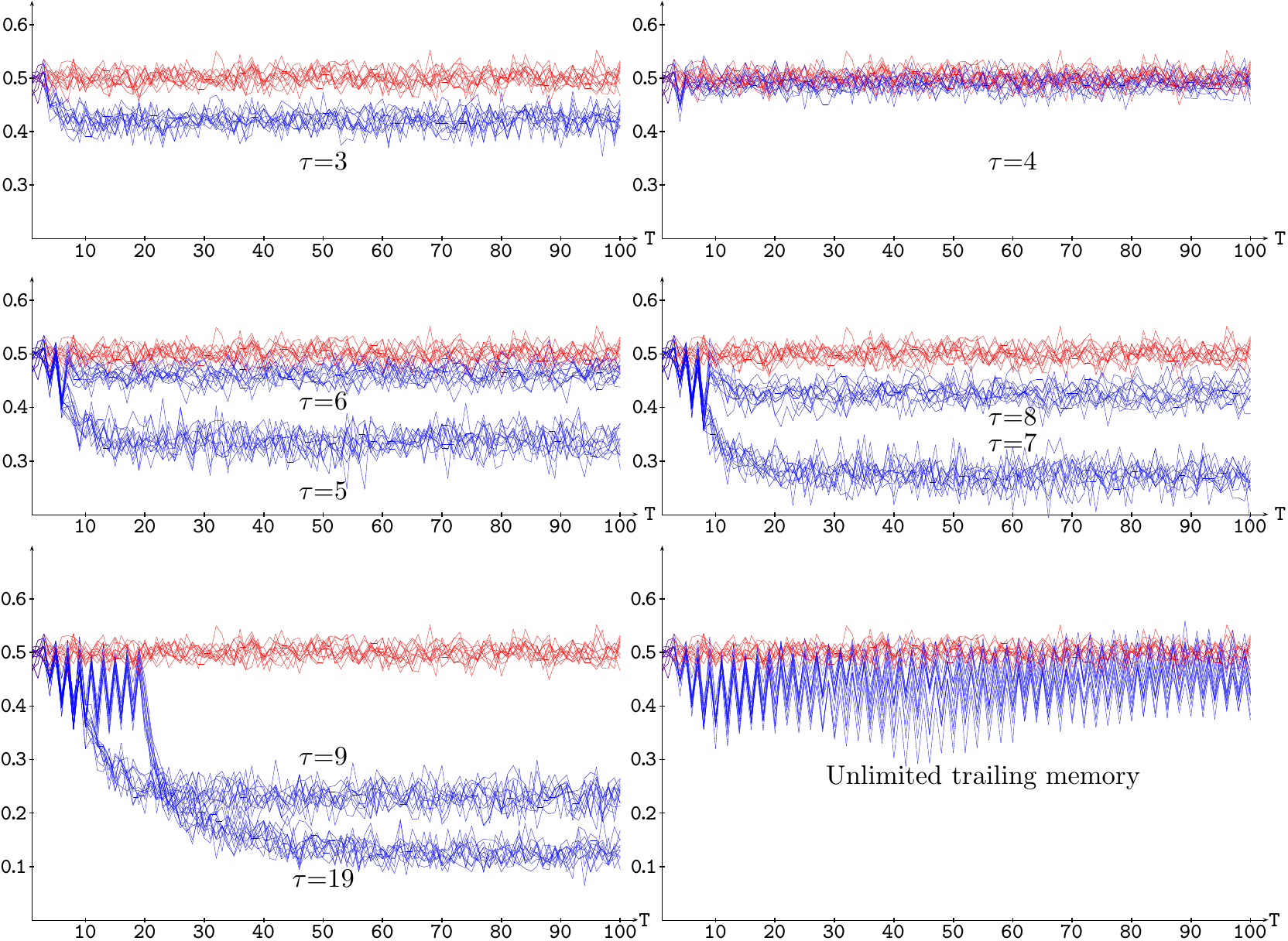}
\caption{Changing rate in eleven simulations up to $T$=100 of the parity rule on  $\beta$=1 skeletons, $N=10^3$\,.
Red plots correspond to the ahistoric simulations, blue plots correspond to $\tau$-majority memory.}\label{fig:crb1}
\end{figure}

\par
Figure~\ref{fig:damageb1} shows the evolution of the damage rate, i.e., the relative Hamming distance between patterns
 resulting from reversing the initial state value of a single node, referred to as damage (or perturbation) spreading.
The damage propagates very rapidly without memory (\textit{butterfly effect}), so that by $T$=30 the red curves already
 oscillate around the fifty per cent of cells. When memory is introduced to the system, the spread of damage is depleted,
albeit the  restraining effect of memory is very low if the charge of memory is low ($\tau$=3,4).
With higher memory lengths,
e.g., $\tau$=9, the depletion in the advance of the damage becomes apparent, though by $T$=150 the damage rate also reaches
the 0.5 level in every simulation in Fig.\,\ref{fig:damageb1}\,. Similar evolution of damage is found with higher limited
trailing memories, such as $\tau$=11 and $\tau$=13 (not shown in the figure). On the contrary case, unlimited trailing memory appears very effective in the control of damage, as shown in the lower-right panel.
\par
Figure~\ref{fig:damageb1} also shows the relative Hamming distance of the ahistoric patterns
 to the corresponding historic ones. After a very short transition period, this distance reaches a fairly permanent plateau level,
oscillating around 0.5 regardless of the charge  of memory endowed.

\begin{figure}[!htb]\centering
\includegraphics[width=1.0\textwidth,draft=false]{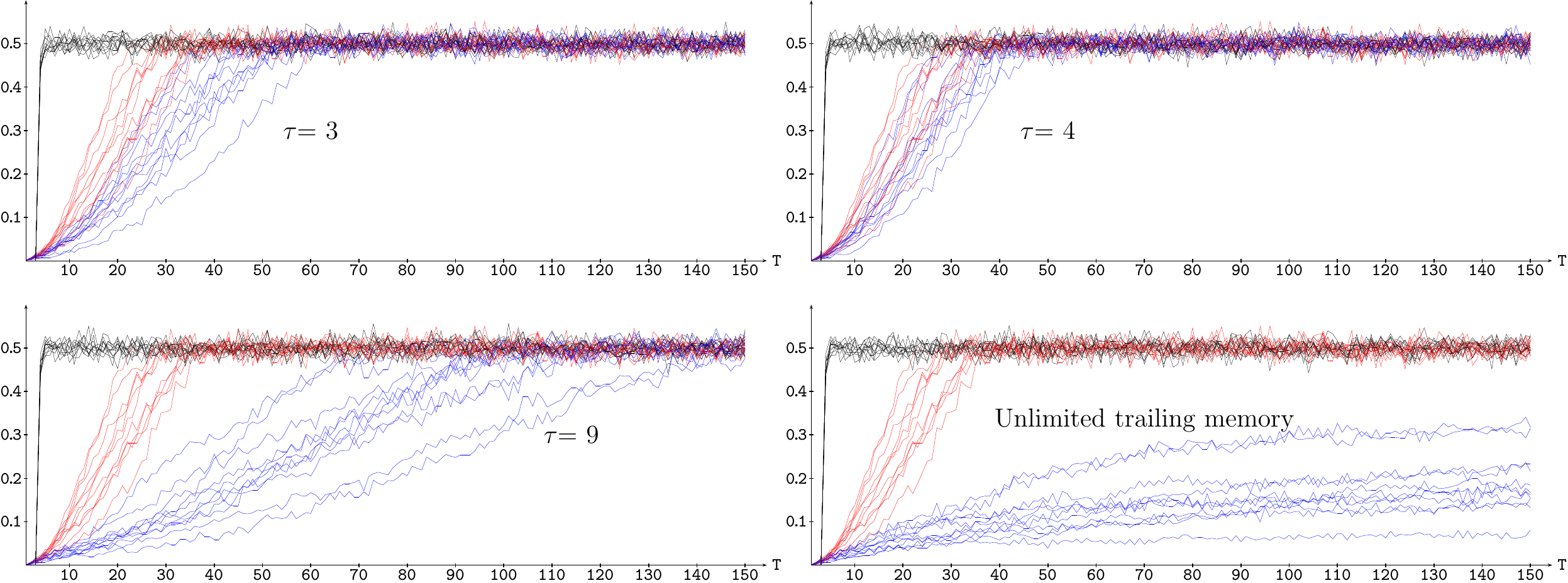}
\caption{Damage spreading in the scenario of Fig.\,\ref{fig:crb1}\,. Color code: black, distance to the ahistoric pattern;
 red, damage spreading with no memory; blue, damage spreading with memory.}
\label{fig:damageb1}
\end{figure}

\pagebreak
\subsection{A $\beta<$1 case.}
\label{betaless1}

Figure\,\ref{fig:exb09} shows a simulation up to $T$=4 of the parity rule on a $\beta$=0.9 skeleton
 based in the same nodes and  initial states as in  Fig.\,\ref{fig:exb1}\,.
In correspondence with the lower $\beta$ value, there are more links connecting nodes in the non-planar graph  of
Fig.\,\ref{fig:exb09} compared to those in Fig.\,\ref{fig:exb1}\,.

\begin{figure}[!htb]\centering
\includegraphics[width=1.0\textwidth,draft=false]{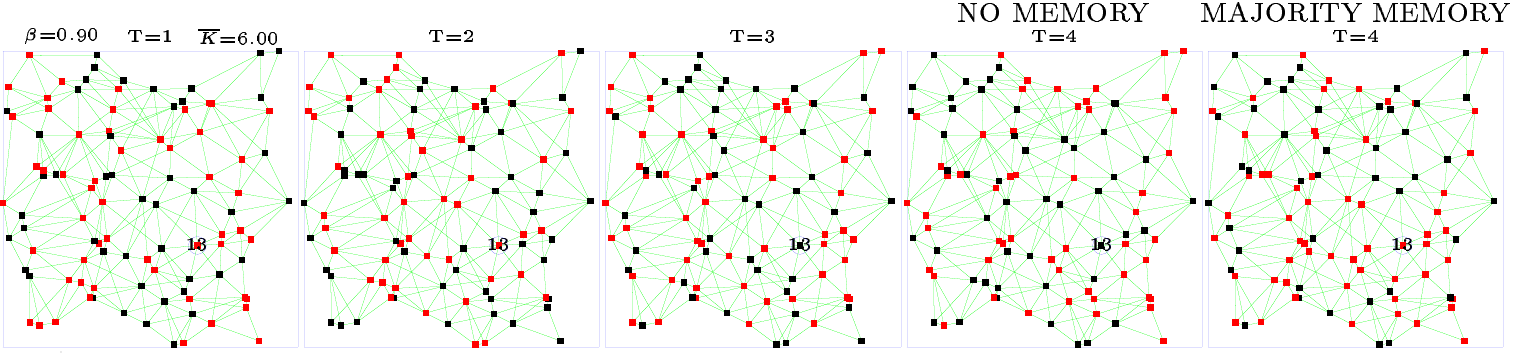}
\caption{A simulation up to $T$=4 of the parity rule on a $\beta$=0.9 skeleton. 
The location of nodes and their initial states are those of Fig.\,\ref{fig:exb1}\,}\label{fig:exb09}
\end{figure}

Figure~\ref{fig:crb09} shows the evolution of the changing rate in eleven different $\beta$=0.9 simulations.
based in the same nodes and initial states as in  Fig.\,\ref{fig:crb1}\,.
 In this scenario, with high connectivity, low memory charges, e.g., $\tau$=3,\,4\,, have not an apparent effect on
 the changing rate, so that there are not presented in the figure. Memory charges of  $\tau$=5
and $\tau$=7 have a limited effect, whereas $\tau$=9,\,19\, already have an apparent effect.
With unlimited trailing memory, the {\it oscillatory} patterns  get a notable amplitude, albeit below 0.5\,.

\begin{figure}[!htb]\centering
\includegraphics[width=1.0\textwidth,draft=false]{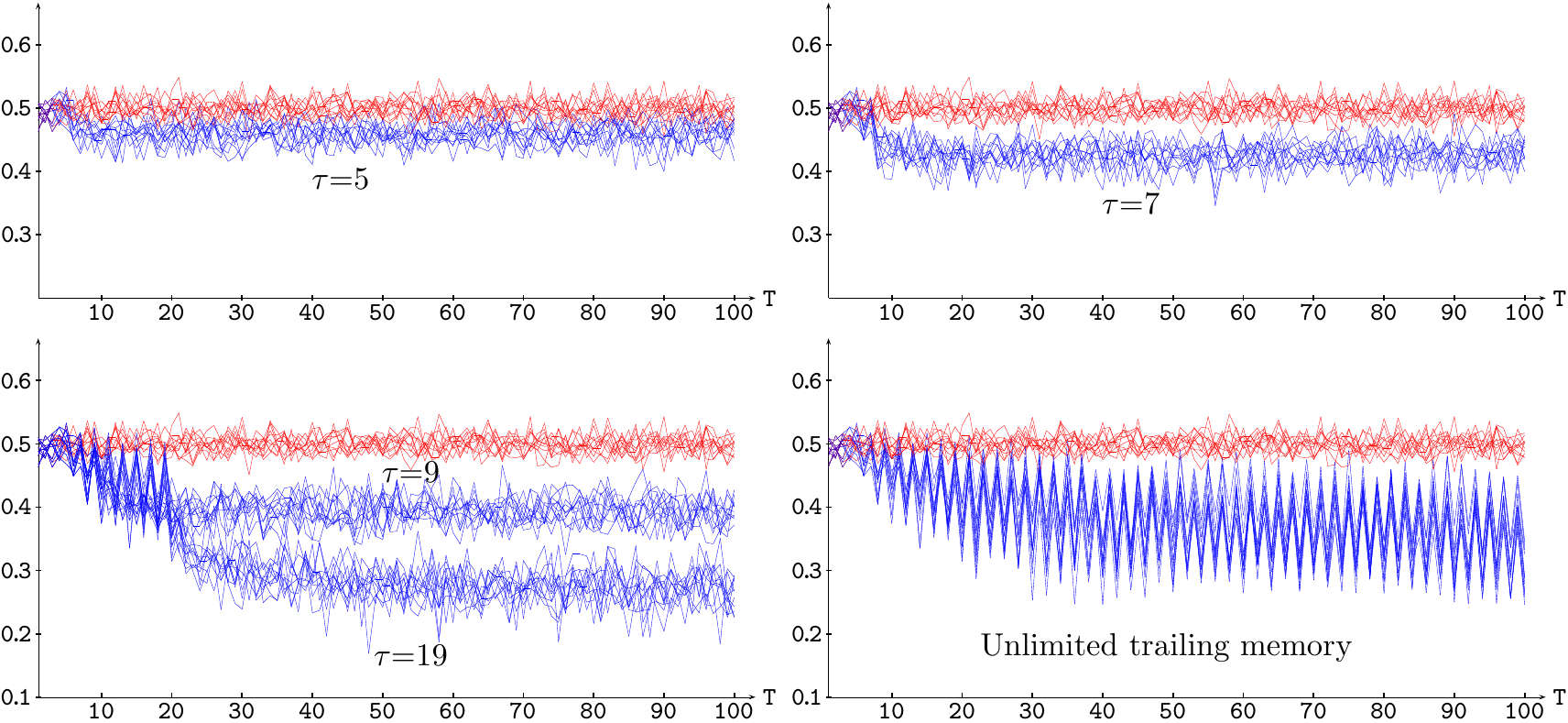}
\caption{Changing rate in eleven simulations up to $T$=100 of the parity rule on  $\beta$=0.9 skeletons. N=1000\,.
Color code as in Fig.\,\ref{fig:crb1}\,.}\label{fig:crb09}
\end{figure}

Figure~\ref{fig:damageb09} shows the damage spreading in the scenario of Fig.\,\ref{fig:crb09}\,.
In this highly connected scenario, memory turns out rather ineffective in the control of damage.
Even with unlimited trailing memory, the damage grows up to 0.5 fairly soon in most simulations,
or is just delayed up to approximately $T$=100 in some others.

\begin{figure}[!htb]\centering
\includegraphics[width=1.0\textwidth,draft=false]{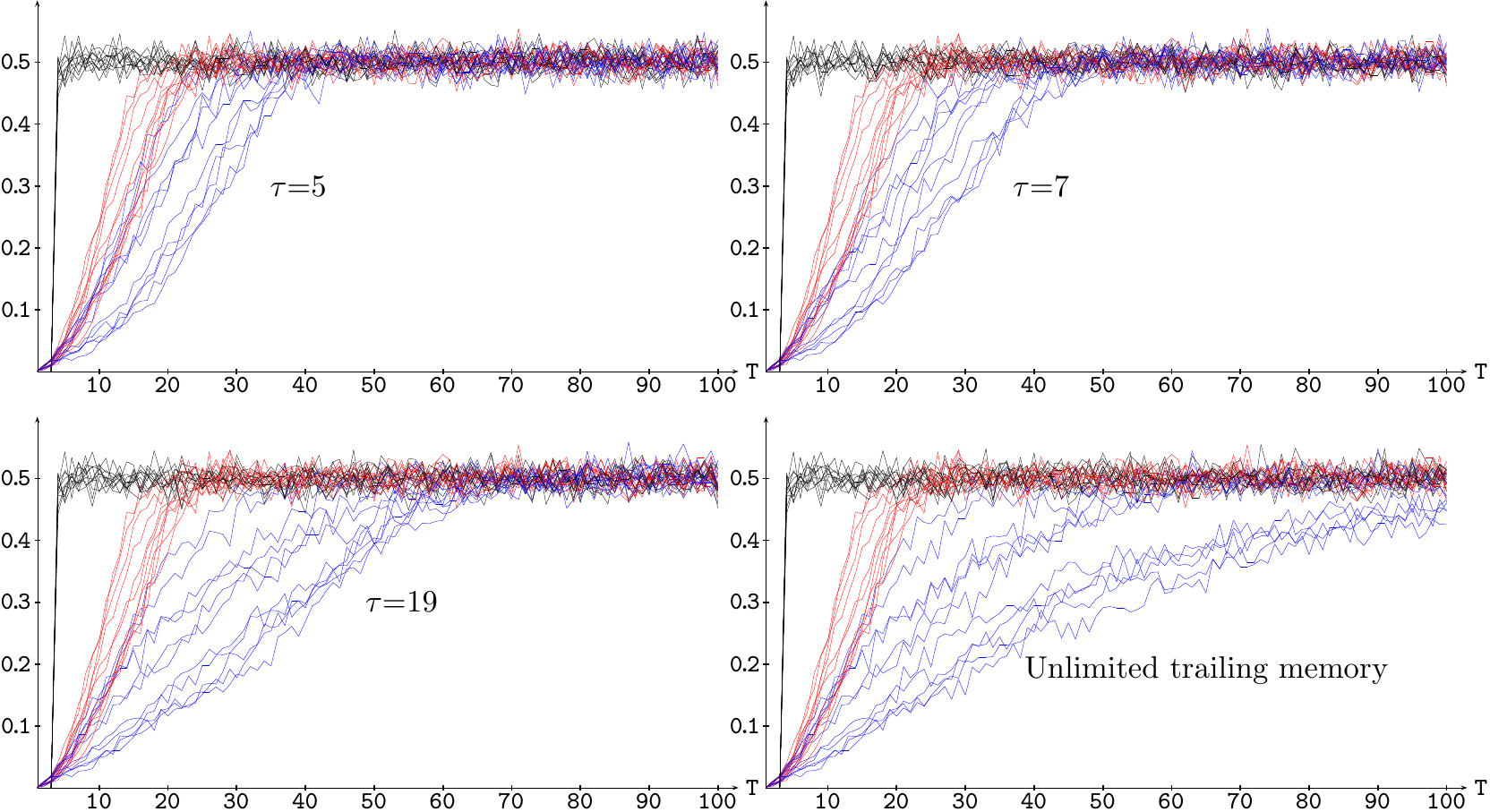}
\caption{Damage spreading in the scenario of Fig.\,\ref{fig:crb09}\,. Color code as in
Fig.\,\ref{fig:damageb1}\,.}\label{fig:damageb09}
\end{figure}

\subsection{A $\beta>$1 case.} \label{betamore1}

Figure~\ref{fig:exb2} shows a simulation up to $T$=4 of the parity rule on a lune-based $\beta$=2 skeleton
(also termed relative neighborhood graphs) based in the same nodes and  initial states as in  Fig.\,\ref{fig:exb1}\,.
In correspondence with the higher $\beta$ value, there are less links connecting nodes in Fig.\,\ref{fig:exb2}
compared to those in Fig.\,\ref{fig:exb1}\,.

\begin{figure}[!htb]\centering
\includegraphics[width=1.0\textwidth,draft=false]{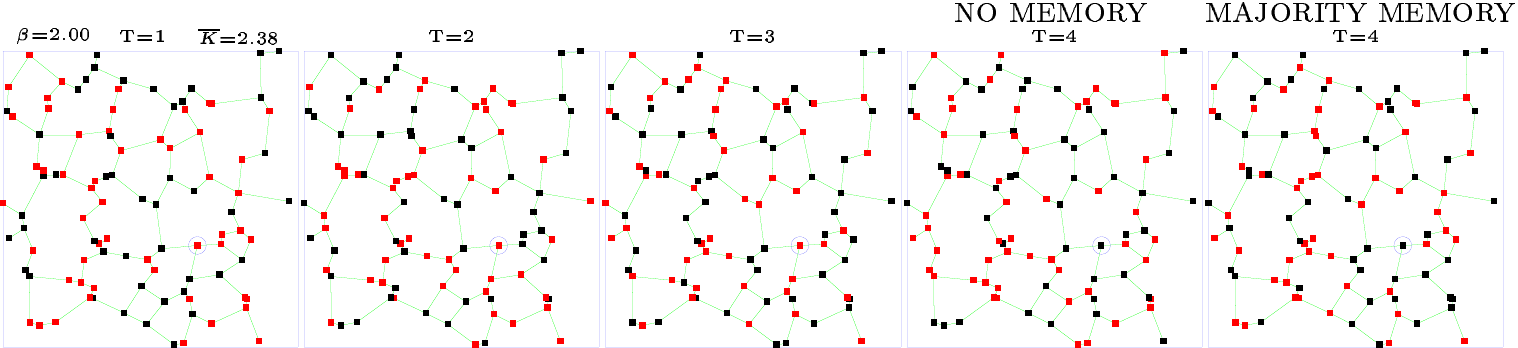}
\caption{A simulation up to $T$=4 of the parity rule on a $\beta$=2 skeleton, $N=10^2$.
The location of nodes and their initial states are those of Fig.\,\ref{fig:exb1}\,.}\label{fig:exb2}
\end{figure}

Figure~\ref{fig:crb2} shows the evolution of the changing rate in eleven different $\beta$=2 simulations
based in the same nodes and initial states as in  Fig.\,\ref{fig:crb1}\,.
 In this scenario, even low memory charges,
e.g., $\tau$=3,\,4\,, have an apparent effect on the changing rate, and higher ones, e.g., $\tau$=9,\,19\,, led this
parameter to very low values. Thus for example with  $\tau$=19\, the changing rate varies in the [0,0.1] interval.
With unlimited trailing memory (lower right panel), the non-truncated {\it oscillatory} patterns soon reach a very high level near 0.5\,.

\begin{figure}[!htb]\centering
\includegraphics[width=1.0\textwidth,draft=false]{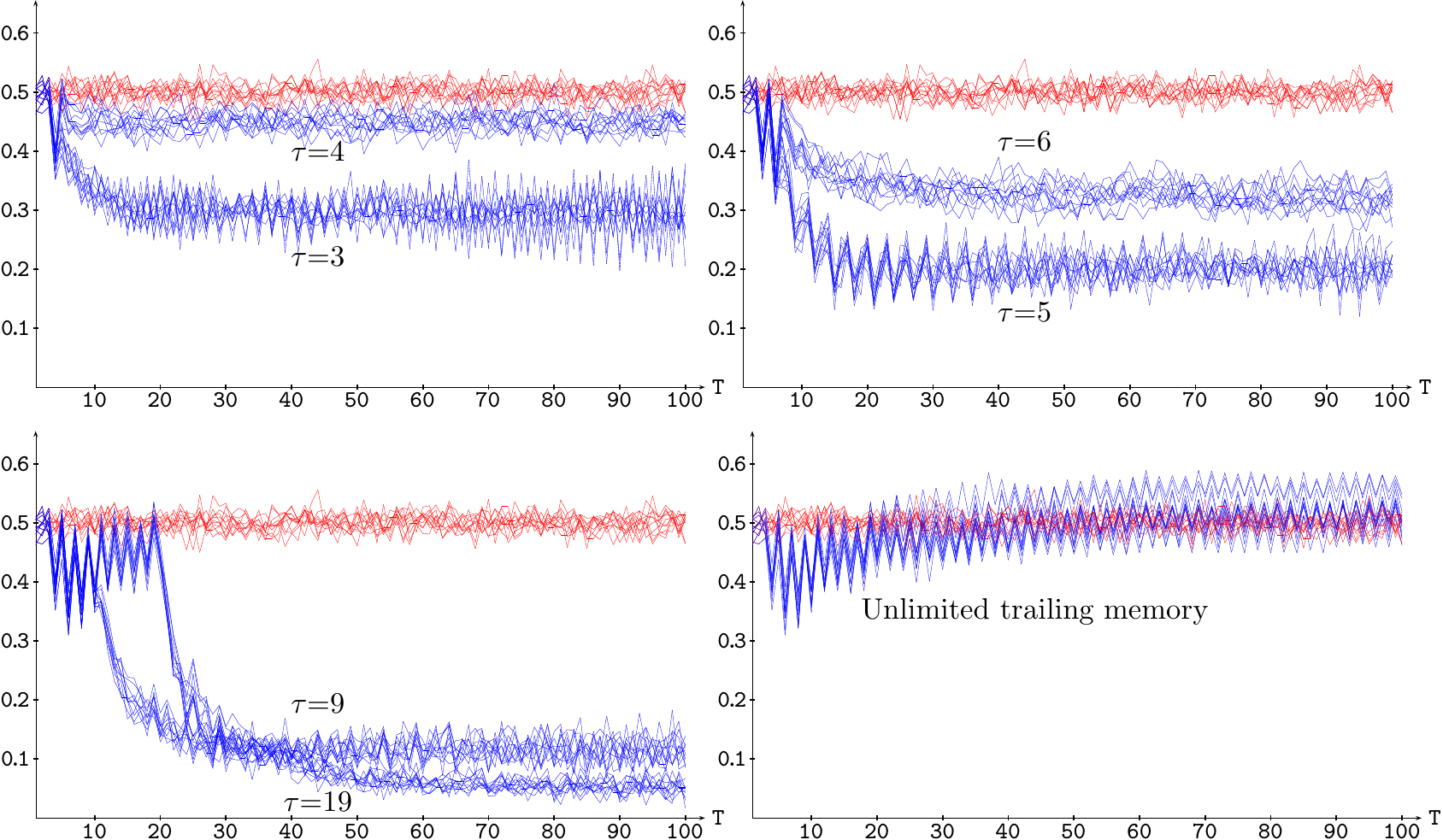}
\caption{Changing rate in eleven simulations up to $T$=100 of the parity rule on  $\beta$=2 skeletons, $N=10^3$\,.
Color code as in Fig.\,\ref{fig:crb1}\,.}\label{fig:crb2}
\end{figure}

\par
Figure~\ref{fig:snapshots} shows the latest snapshots of the changing nodes in one of the $\beta$=2 skeleton simulations
 in Fig.\,\ref{fig:crb2}\,. Both in the ahistoric context, where approximately half of the nodes change,  
and with $\tau$=19 majority memory, in which case only small sets of nodes are changing.

\begin{figure}[!htb]\centering
\includegraphics[width=1.0\textwidth,draft=false]{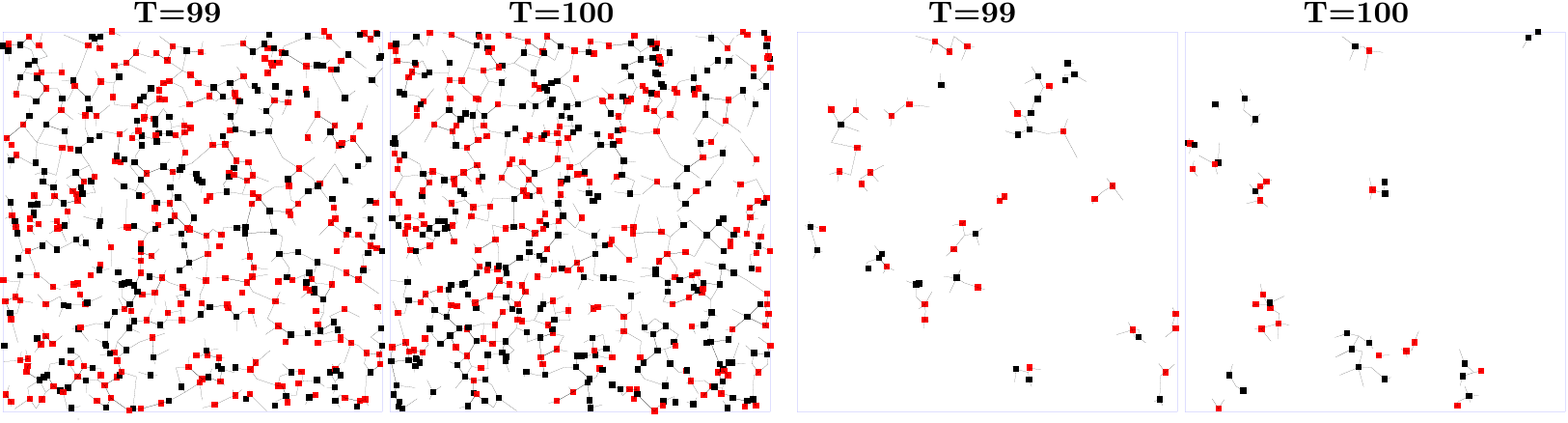}
\caption{Latest snapshots of the changing nodes in one of the $\beta$=2 skeleton simulations in Fig.\,\ref{fig:crb2}\,.
Color code: Red\,: 0$\rightarrow$1\,, Black\,: 1$\rightarrow$0\,. Ahistoric patterns (left), and patterns with 
 $\tau$=19 majority memory (right)\,.}\label{fig:snapshots}
\end{figure}

\par

Figure~\ref{fig:damageb2} shows the damage spreading in the scenario of Fig.\,\ref{fig:crb2}\,.
In this scenario, with low connectivity, the depletion in the advance of the damage becomes apparent already
with $\tau$=3 though by $T$=100 the damage rate also reaches the 0.5 level. As expected, $\tau$=9 is much more
effective in the damage control, and, above all, unlimited trailing memory which virtually impedes the advance
of damage.

\begin{figure}[!htb]\centering
\includegraphics[width=1.0\textwidth,draft=false]{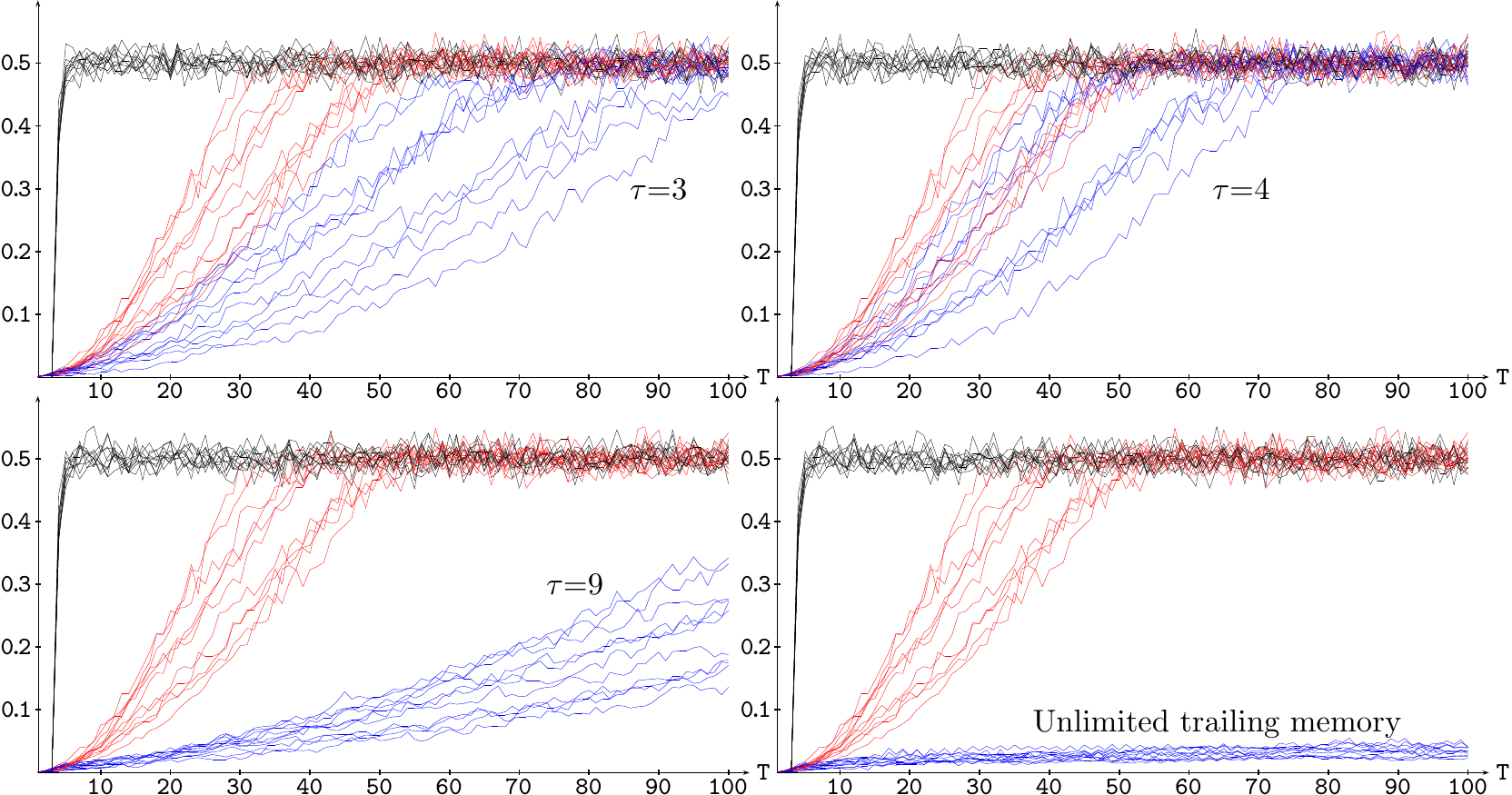}
\caption{Damage spreading in the scenario of Fig.\,\ref{fig:crb2}\,. Color code as in Fig.\,\ref{fig:damageb1}\,.}
\label{fig:damageb2}
\end{figure}

\section{Asymptotic levels.}\label{overview}

Figure \ref{fig:overview} shows the asymptotic changing rate (left) and damage spreading (right)
with increasing values of the length of memory $\tau$ in the three $\beta$-scenarios considered before.
 {\it Asymptotic} standing for the mean of then 
ten last values in both cases.  The scenario is that of previous sections, i.e., a $N$=1000 network
 run up to $T=100$ iterations, so that $\tau$=100 implies full memory.  The changing rate notably declines 
as a result of the initial increase of the length of memory (particularly with $\beta$=0.9\,). The lowest
changing rate levels are achieved around $\tau \simeq T/2$=50. These levels are not very much altered
until a very high length of memory is reached. Namely, by $\tau \gtrsim 9T/10$=90, the oscillation in the
changing rate is maintained most of the time-steps, so that the subsequent decay becomes increasingly difficult. 

\begin{figure}[!htb]\centering
\includegraphics[width=1.0\textwidth,draft=false]{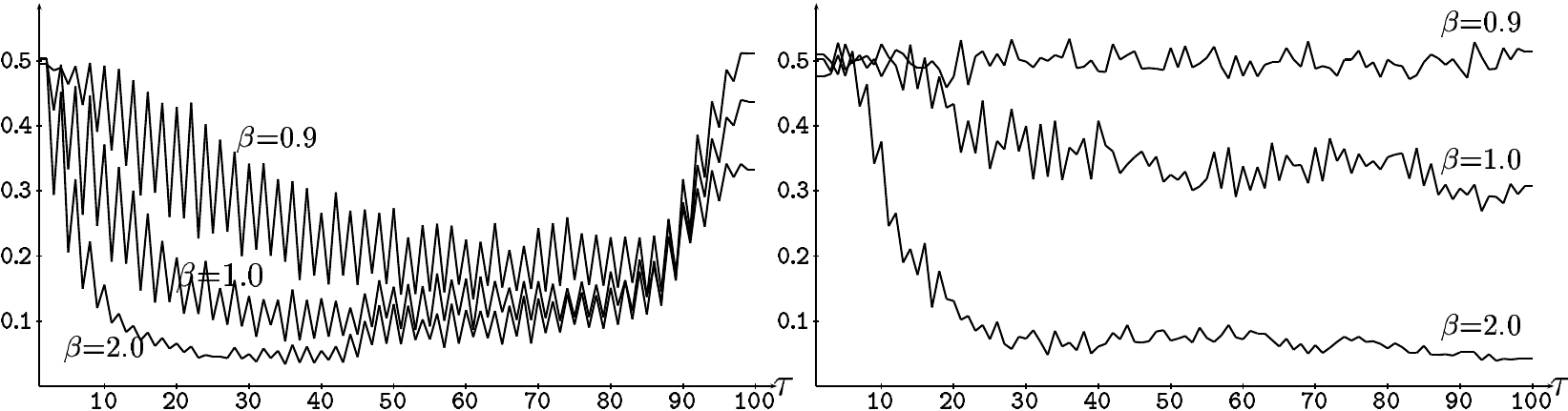}
\caption{Asymptotic changing rate (left) and damage spreading (right) in $N$=1000 
parity $\beta$-skeletons with $\tau$-majority memory.}
\label{fig:overview}
\end{figure}

The effect of increasing the memory length on the damage spreading (Fig.\,\ref{fig:overview}, right), is much 
more monotone\,: a seemingly plateau, below the 0.5 level, is reached in the $\beta$=1.0 and  $\beta$=2.0  
cases, whereas in the highly connected $\beta$=0.9 context, memory does not qualitatively alter the maximum 
0.5 level.

The  effect of memory in bigger networks, does not qualitatively differ from that presented here for $N$=1000.
The dynamics of both the changing rate and damage spreading behave in broad strokes as in the $N$=1000 case. Maybe in a smoother
way in every single simulation, and closer when considering different simulations. But smaller networks are more problematic.   
Thus, Fig.\ref{fig:overviewN100} shows the asymptotic changing rate (left) and damage spreading (right)
in the same conditions of Fig.\,\ref{fig:overview}, except in what refers to the network size\,:
$N=100$ instead of $N=1000$\,. The general pattern of the asymptotic changing rate frame in Fig.\,\ref{fig:overviewN100}
 resembles that of Fig.\,\ref{fig:overview}, albeit the higher variability of the changing rate in the small network 
is hidden by the mere presentation of the average of the last values. The structure of the damage spreading frame 
changes dramatically compared to that of the bigger network\,: memory is unable to avoid the full damage spread across small networks.
 Neither in the low connected scenarios, not even in that with $\beta$=2.0\,. 

\begin{figure}[!htb]\centering
\includegraphics[width=1.0\textwidth,draft=false]{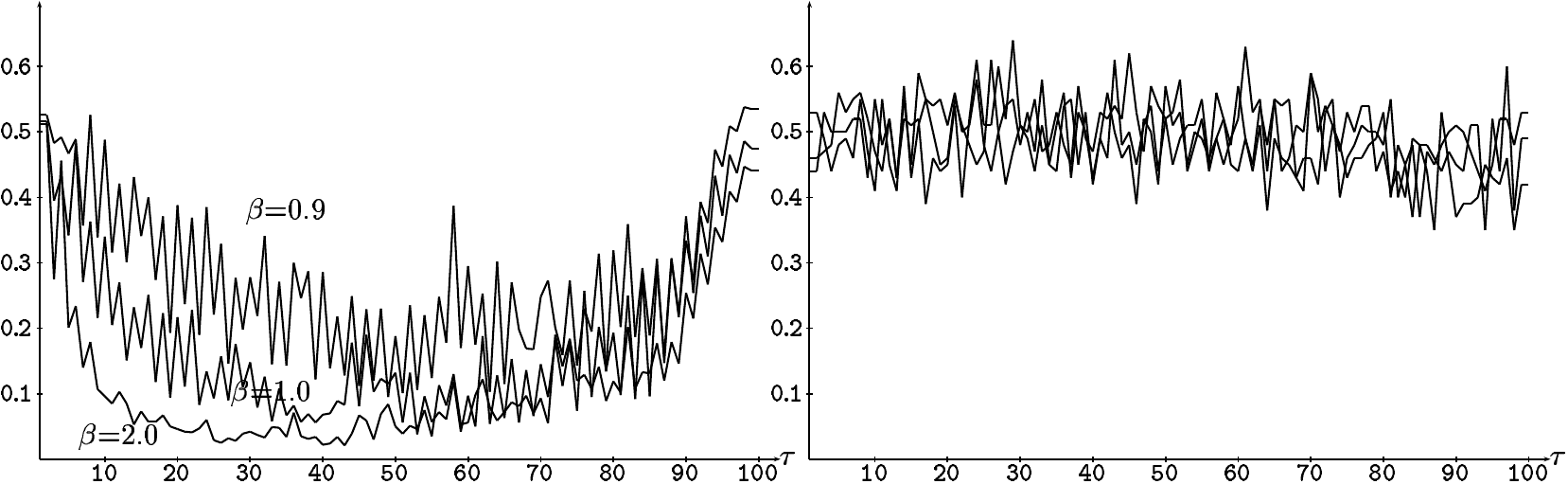}
\caption{Asymptotic changing rate (left) and damage spreading (right) in $N$=100 parity $\beta$-skeletons with $\tau$-majority memory.}
\label{fig:overviewN100}
\end{figure}

\section{Weighted Memory.}\label{weightedmemory}

Majority memory of the last $\tau$ time-steps demands $\tau$ extra bits per node to store
their corresponding state values. To avoid this drawback, past state values can be weighted in such a
way that only the accumulated memory charge needs to be stored.
Thus, for example, historic memory can be weighted by applying a geometric discounting process
in which the state $\sigma ^{(T-t)}_{i}$, obtained $t$ time steps
before the last round, is actualized to  $\alpha ^{t}\sigma
^{(T-t)}_{i}$, $\alpha$ being the \textit{memory factor} lying in the [0,1]
interval. This well known mechanism fully takes into account the last round $(\alpha
^{0}=1)$, and tends to \textit{forget} the older rounds.
\par
	Every node will be featured by the rounded weighted mean of all its past
states, so the memory  mechanism is implemented in two steps at time-step $T$\,:
\par
$(i)$ The unrounded weighted mean ($m$) of the states of every node is computed first\,:\vspace{-.15cm}
\[
m^{(T)}_{i}(\sigma ^{(1)}_{i},\ldots ,\sigma ^{(T)}_{i}) =
{\displaystyle\frac{\sigma^{(T)}_{i} + \displaystyle \sum^{T-1}_{t=1}\alpha ^{T-t}\sigma
^{(t)}_{i}}{1+ \displaystyle \sum^{T-1}_{t=1}\alpha ^{T-t}}}
\equiv {\displaystyle\frac{{{\omega^{(T)}_{i}}} }{\Omega(T) }}
\] \label{eq:meq}

$(ii)$ Then the trait state \textit{s} is obtained by
comparing the memory $charge$ $m$ to the landmark  0.5  (if $\sigma\in\{0,1\}$), assigning the last state in case of an equality to this value, so that\,:\vspace{-.25cm}
\[
s^{(T)}_{i} ={\cal H}(m_{i}^{(T)})=\left \{
\begin{array}{lll}
1 & ~if~~& m^{(T)}_{i} > 0.5 \\
\sigma^{(T)}_{i} & ~if~~ & m^{(T)}_{i} = 0.5 \\
0 & ~if~~ & m^{(T)}_{i} < 0.5 ~~~.
\end{array}\right .\]
\par
The choice of the memory factor $\alpha$ simulates the long-term or remnant memory effect: the limit case
 $\alpha =1$ corresponds to a memory with equally weighted records
($full$ memory, equivalent to unlimited trailing \textit{majority} memory), whereas $\alpha \ll 1$ intensifies the contribution
 of the most recent states and diminishes the contribution of the more remote states (short-term working
 memory). The choice $\alpha = 0$ leads to the ahistoric model. This memory implementation will be referred to as
 $\alpha$-memory.
\par
This geometric memory mechanism is not  \textit{holistic} but
\textit{cumulative} in its demand for knowledge of past history:
the whole $\big\{ {\sigma}^{(t)}_{i}\big\}$ series needs not be known
to calculate the term $\omega^{(T)}_{i}$ of the memory \textit{charge} $m^{(T)}_{i}$,
while to (sequentially) calculate $\omega^{(T)}_{i}$ one can
resort to the already calculated $\omega ^{(T-1)}_{i}$ and compute: $\omega
^{(T)}_{i}= \alpha \omega ^{(T-1)}_{i}+\sigma ^{(T)}_{i}$. Consequently, only one
 number per node ($\omega_i$) needs to be stored. This positive property is accompanied by the drawback
of any weighted average memory:\, it computes with real numbers instead of with integers as is done
with majority memory.
\par
In the most unbalanced scenario,
 $\sigma^{(1)}_{i}= ... =\sigma ^{(T-1)}_{i} \neq \sigma ^{(T)}_{i}$, it holds that\,\footnote{
\vspace{-0.05cm}
\[\left \{
 \begin{array}{lll}
 m(0,0,\ldots,0,1)=\displaystyle\frac{1}{2}~\quad\equiv~\quad$\quad$~1~~$\quad$=
  \displaystyle\frac{1}{2}\displaystyle\frac{\alpha^{T}-1}{\alpha-1}
 \\ \\
 m(1,1,\ldots,1,0)=\displaystyle\frac{1}{2}~\quad\equiv~\quad
  \displaystyle\frac{\alpha^{T}-1}{\alpha-1}=\displaystyle\frac{1}{2}\displaystyle\frac{\alpha^{T}-1}{\alpha-1}
 ~~~
\end{array} \right .
\Rightarrow \alpha^{T}_{T}-2\alpha_{T}+1=0  \]}\,: $m = 0.5
\Rightarrow\alpha^{T}_{T}-2\alpha_{T}+1=0$ \label{eq:eq1} [\ref{eq:eq1}], where
 $\alpha_T$  holds for the critical value of $\alpha$ below which memory has no effect up to time-step $T$\,.
When $T\rightarrow \infty $, the equation $[\ref{eq:eq1}]$  becomes\,: $-2\alpha_{\infty}+1=0$, thus,
 $\alpha$-memory is not effective if $\alpha \le 0.5$\,.
\par

\par
Figure~\ref{fig:cralpha} shows the changing rate with $\alpha$-memory in parity $\beta$-skeletons of one
thousand nodes, starting from a single active node. Without memory, the changing rate progresses very fast regardless
of the $\beta$ value. On the contrary, full memory, i.e., $\alpha=$1.0, dramatically restrains the changing rate.
In the $\beta$=2.0 scenario, thus with low connectivity, even the smallest memory charges, e.g., $\alpha$=0.6, lead
the changing rate to extinction. In the more connected networks, low memories such as  $\alpha$=0.6 and $\alpha$=0.7
only delay the increase of the changing rate, that reaches 0.5 in the figure in both the $\beta$=1.0 and $\beta$=0.9
scenarios. With $\alpha$=0.8 memory, the 0.5 level is not reached up to $T$=100 in the $\beta$=1.0 simulation,
but it is reached in the $\beta$=0.9 simulation. Finally, with $\alpha$=0.9 the changing rate is notably restrained in
both scenarios, though it seems not stabilized as it is with full memory. Thus, in the simple context of
Fig.\,\ref{fig:cralpha} full memory exerts a, let us say, expected effect, that of maximum dynamic inhibition. The dynamic
starting from a single active node may be considered as the simplest form of damage spreading, measured by the evolution of 
density. The density curves (not shown) mimic those of the changing rate shown in Fig.\,\ref{fig:cralpha}\,, including those with
 full memory, agreeing with the expected effect on damage spreading already shown in the majority memory scenario.  

\begin{figure}[!htb]\centering
\includegraphics[width=1.0\textwidth,draft=false]{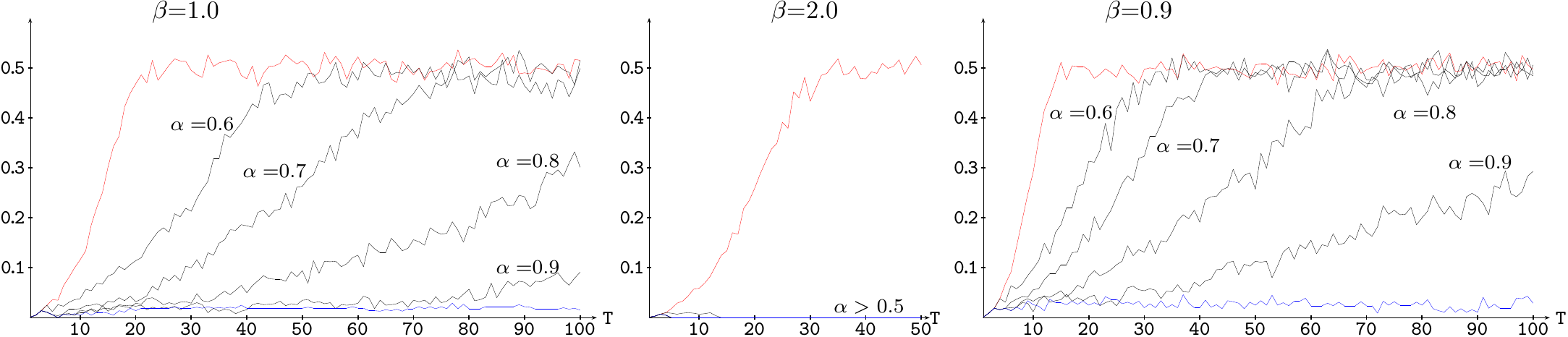}
\caption{Changing rate with $\alpha$-memory in parity $\beta$-skeletons from a single active node. N=1000\,.
Red plots correspond to the ahistoric simulations, blue plots correspond to full majority memory.}\label{fig:cralpha}
\end{figure}

\par

Figure \ref{fig:overviewalpha} shows the asymptotic changing rate (left) and damage spreading (right) in $N$=1000 
parity $\beta$-skeletons with $\alpha$-memory. The effect of $\alpha$-memory is much more monotone than that of 
majority memory shown in Figure \ref{fig:overview}\,. Thus, as a rule, increasing the memory factor implies restrain in both
the changing rate and the damage spreading. The changing rate in Fig.\,\ref{fig:overviewalpha} sharply peaks in the full memory 
($\alpha$=1.0) scenario, according to the the full lengt ($\tau$=100) majority memory changing rates shown in Fig.\,\ref{fig:overview}\,.   
It turns out remarkable that damage spreading is restrained only when $\alpha$ reaches high values\,:
 $\gtrsim 0.90$ in the medium connected $\beta$=1.0 scenario, and   
 $\gtrsim 0.85$ in the low connected $\beta$=2.0 scenario.  

\begin{figure}[!htb]\centering
\includegraphics[width=1.0\textwidth,draft=false]{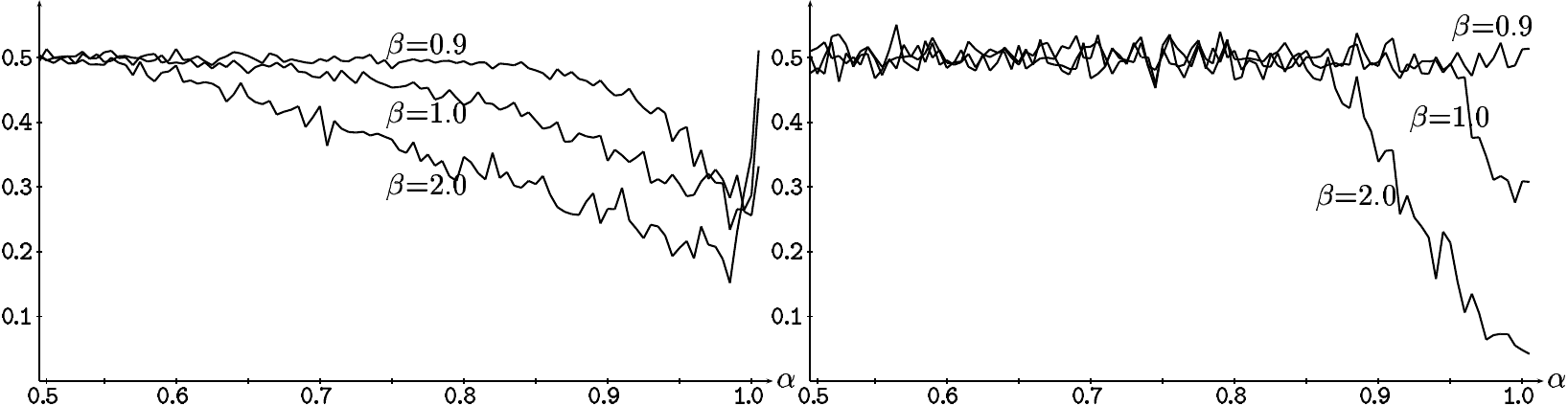}
\caption{Asymptotic changing rate (left) and damage spreading (right) in $N$=1000 
parity $\beta$-skeletons with $\alpha$-memory.}
\label{fig:overviewalpha}
\end{figure}

\section{Other rules, other contexts}\label{otherrules}

The results presented up to now deal with the {\it parity} rule. This rule is a particularly `active' rule 
that produces very high changing rates. Nevertheless, most other totalistic rules, i.e., based in the sum of the
neighbour states $\small{\sum_i} =\displaystyle\sum_{j\in\mathcal{N}_{i}}\sigma^{(T)}_{j}$\,,
tend to led the dynamic to low stable levels of changing rate. 
This feature is particularly visible in {\it majority} rule\,:
$\sigma^{(T+1)}_{i}$=1 if $2\sum_i  > K_i\;$, $\sigma^{(T+1)}_{i}$=0, if $2\sum_i > K_i\;$ 
where $K_i $ standing for the connectivity of node $i$\,.

\begin{figure}[!htb]\centering
\includegraphics[width=1.0\textwidth,draft=false]{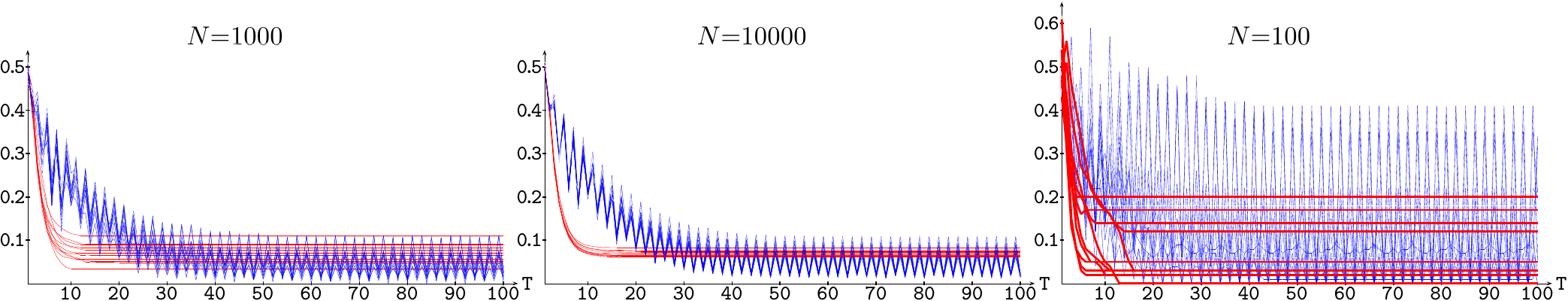}
\caption{Changing rate with the ahistoric majority rule (red), and the
 majority rule  with nodes endowed memory of the  parity  of the last three states (blue). 
 $\beta=$1-skeletons with $10^3$ nodes (left), $10^4$ nodes (center), and $10^2$ nodes (right). }
\label{fig:memoparity} 
\end{figure}

\par
In the simulations presented in Fig.~\ref{fig:memoparity}, the generic node $i$ preserves its
last state in case of a tie: $\sigma^{(T+1)}_{i}$=$\sigma^{(T)}_{i}$ if $2\sum_i = K_i$\,. 
The dynamics is very much
sensitive at this respect, so that adopting  $\sigma^{(T+1)}_{i}$=1 if $2\sum_i = K_i$ will lead the density to 1.0,
whereas with $\sigma^{(T+1)}_{i}$=0 if $2\sum_i = K_i$ the density will tend to plummet. With the {\it neutral} criterion 
adopted here, there is no reason for density drift to any of the extreme values, so that the red lines in 
Fig.~\ref{fig:memoparity} become stabilized after a short transition period. 
\par
In this scenario, the activity may be supported implementing memory rules not of majority type,
 such as the parity rule acting as memory\,\cite{RAS22}\,. Thus for example, the parity of
the last three states\,\footnote{\hspace{0.25cm} $s^{(1)}_i=\sigma^{(1)}_i\;,$~
$s^{(2)}_i=\sigma^{(1)}_i \oplus \sigma^{(2)}_i\;$,~
$s^{(T)}_i=\sigma^{(T-2)}_i \oplus \sigma^{(T-1)}_i \oplus \sigma^{(T)}_i\,$.}, as implemented in Fig.\,\ref{fig:memoparity}\,.
The left panel of this figure applies to simulations with $N=10^3$ nodes (usually for present paper). 
Simulations with higher number of nodes, e.g., $N=10^4$ in the central panel, 
do not qualitatively differ from those with $N=10^3$\,. They present the same general features in their dynamics,
 maybe with an additional characteristic\,:
the evolution of density, changing rate and damage spreading is much more similar among them. Opposite, simulations with few nodes,
e.g., one hundred nodes in the right panel, are much more unforeseeable. Not only in what respect to the effect of memory, even
the ahistoric dynamics may converge to very different densities and changing rates. Thus, for example, in the $N=10^2$ simulations
of the far right panel of Fig.\,\ref{fig:memoparity}\,, some ahistoric dynamics are led  to very low densities 
of $\sigma$=1 nodes, e.g., $\rho$=0.2, whereas other ones reach high densities, e.g., $\rho$=0.8, very distant from the $\rho$=0.5 
obtained in ahistoric simulations with $N=10^3$ or $N=10^4$ nodes.
\par
\par
We have studied the effect of memory  in several cellular automata and  Boolean network
 scenarios\,\cite{RASBULL,RASCAR1,RASCAR2}\,. The mean connectivity across the eleven simulations run here is $\overline{K}$=3.85,
 $\overline{K}$=2.50, and  $\overline{K}$=6.64, for $\beta$=1.0,  $\beta$=2.0, and  $\beta$=0.9
respectively. The dynamics studied in this article may be somehow compared to those 
 simulations in the referenced articles with mean connectivities approaching these values.
Thus, for example, the way in which a single active node propagates activity (a kind of damage 
spreading), as shown in the patterns at $T$=100 in Fig.\,\ref{fig:paternscrb1}\,:  no restrain in the ahistoric simulation,
some dispersion with $\alpha$=0.9, and activity concentrated near the initial active node with full memory.     

\begin{figure}[!htb]\centering
\includegraphics[width=1.0\textwidth,draft=false]{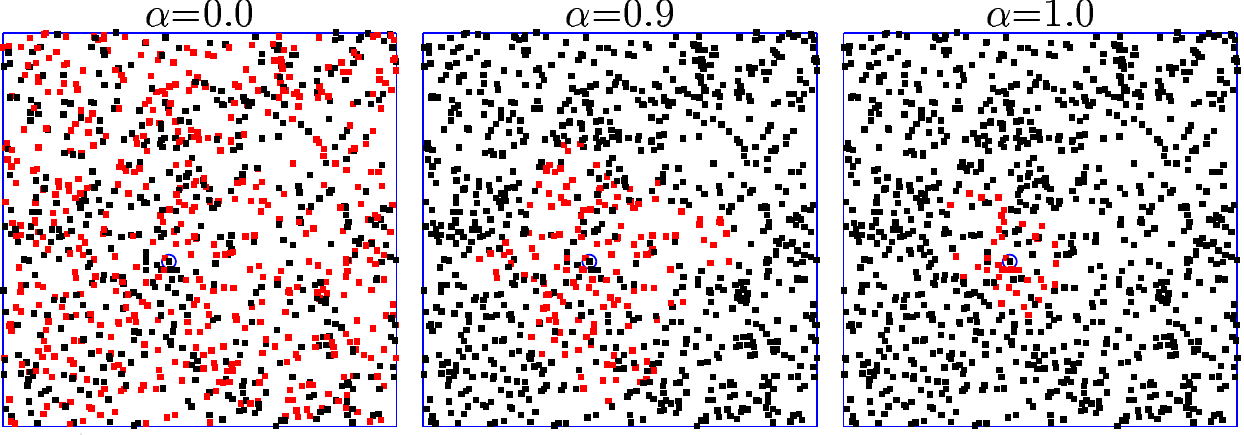}
\caption{Patterns at $T$=100 with $\alpha$-memory in the parity $\beta$=1-skeleton starting from a single active node
considered in Fig.\,\ref{fig:cralpha}\,.}\label{fig:paternscrb1}
\end{figure}

\par
In previous studies we have analyzed the effect of memory in some spatialized games. Thus, the battle of the sexes\,\cite{RASBOS} and the 
prisoner's dilemma (PD)\,\cite{RASCHAOS,RASBIOS}\,.
It has been concluded in those studies that the consideration of previous choices and payoffs in the  updating rule, notably boosts
cooperation. Although the effect of memory on games with players located in $\beta$-skeleton nodes is planned for further work, let us
point here that the just outlined conclusion, reached in other spatial contexts, may likely also be found when dealing with 
the PD $\beta$-skeletons. Thus for example, starting with a single defector in a sea of cooperators, memory will restrain the defection 
progress, much in the way of the restrain of the advance of the active nodes in Fig.\,\ref{fig:paternscrb1}\,. Memory will likely have
even a stronger effect on  the PD dynamics in $\beta$-skeletons with low connectivities, i.e., large $\beta$. On the contrary, in 
highly connected $\beta$-skeletons, i.e., small $\beta$, it is  likely to be found that memory will be unable
 to avoid the spread of defection
propelled via action at distance. 

\section{Potential applications}\label{applications}

What are potential outcomes of our findings\,?.  Proximity graphs, particularly relative neighbourhood graphs ($\beta$-skeletons for
$\beta=2$) are invaluable in simulation of human-made road networks, see for example the  study of Tsukuba central
district~\cite{watanabe_2005, watanabe_2008}. The graphs provide a good formal representation of biological
transport networks, particularly foraging trails of ants~\cite{adamatzky_2002} and protoplasmic networks of slime mold \emph{Physarum polycephalum}~\cite{adamatzky_ppl_2008,adamatzky_jones_2009}. Our computational approaches could be applied
toward studies of damage propagation in artificial and natural networks, and also used to develop novel techniques for
controlling space-time dynamics on the networks.

Majority of the results in the paper deal with parity rule. This rule is important because parity rule cellular automata are considered to be low level models of self-replication, see e.g.~\cite{sipper_1998, Julian}, and also, similar rules are used for 
fast simulation of soliton cellular automata, see e.g.~\cite{park_1998}. Ultimately the parity rule can be considered as a good discrete approximation of propagation of perturbations in \emph{active} non-linear discrete media, where a localised traveling 
perturbation (analog of excitation in excitable media) generates other perturbations. Thus, results of the paper could be, in principle, applied to studying of memory-based control of spatio-temporal dynamics of non-linear active media.

With regards to smile mold, indeed its protoplasmic network is always changing, new tubes are formed, some old tubes become abandoned. Even in this case --- \emph{`a la structure dans le flux'} --- $\beta$-skeleton is an excellent approximation of protoplasmic networks for over 80\% of plasmodium foraging span~\cite{AdamatzkyPhysarumMachines}. The parity simulated in the paper generates non-trivial patterns of activity, which can be interpreted in terms of oscillatorty dynamics of electrical potential or contractility of the plasmodium of \emph{Physarum polycephalum}~\cite{tsuda_jones}. By tuning capacity of a node memory of $\beta$-skeleton we could shape periodical activity of the graph and thus imitate interactions between electrical and oscillatory 
activity of the slime mould, and also incorporate heterogeneous network of discrete bio-chemical oscillators in the discrete models of the plasmodium~\cite{adamatzky_naturewissenschaften_2007}. 

Yet another, rather more specific domain where our results on $\beta$-skeleton with memory can be applied is a computation in assembles of lipid vesicles filled with Belousov-Zhabotinsky (BZ) mixture~\cite{gorecki_private,neuneu}. When implementing computation in excitable chemical systems~\cite{AdamatzkyRDC} we usually encounter an a troublesome problem of wave-fragments' instability. A traveling wave-fragment (which represents a value of logical variable) rarely preserves its shape for a long time, it rather collapses or expands. This problem can be overcome by subdividing computing substrate into interconnected compartments --- BZ-vesicles --- and allowing waves to collide one with another only inside the compartments, see details in~\cite{adamatzky_BZvesicles, holley_2010}. An ensemble of BZ-vesicles forms a microscopic analog of a massively-parallel processor. In computer simulations~\cite{adamatzky_BZvesicles, holley_2010} BZ-vesicles are arranged in regular arrays.  It is very unlikely that in natural conditions ensembles of BZ-vesicles will be regular.  When BZ-vesicles, quite possible of different sizes, are aggregated into an assemble and tightly packed, they are represented by Delaunay triangulation. A computation on Delaunay
triangulation is implemented in the same manner as on a regular cellular automata~\cite{adamatzky_delaunay}. However, when some BZ-vesicles start to degrade, and other increase in size average connectivity of the vesicle conglomerate decreases. What kind of proximity graph represents then the ensemble of BZ-vesicles?  

In 1980 Toussaint demonstrated that MST$\subseteq$RNG$\subseteq$DT~\cite{toussaint_1980}.   This hierarchy of graphs was later enriched with GG~\cite{matula_1980} as follows:  

\vspace{0.2cm}

\centerline{
NNG $\subseteq$ MST $\subseteq$ RNG $\subseteq$ GG $\subseteq$ DT
}

\vspace{0.2cm}

That is when BZ-vesicles bursting the graph representation of the vesicles' ensemble will travel downwards in the Toussaint hierarchy: the diagram becomes a Gabriel graph GG ($\beta$-skeleton for $\beta=1$) then relative neighbourhood graph RNG
($\beta$-skeleton for $\beta=2$) then spanning tree MST and then disconnected nearest-neighbour graph. Response of each BZ-vesicle to the excitation in the neighbouring vesicles depends on the diffusion speed, size of pores between the vesicles, concentration of 
reactants inside the vesicle and overall age of the system. Usually, vesicle's reaction in any particular time step 
is determined not only by current dynamics of excitation in the vesicle's neighbourhood but also by history of chemical reactions in the vesicle itself. Thus, in computational experiments with $\beta$-skeletons with memory we discover, at least at the level of discrete yet fine-grained models, how activity can propagate in a ensemble of `ageing' BZ-vesicles and what methods of the activity control can be employed.

\section{Conclusion}\label{conclusion}

As a rule, endowing nodes in beta-skeleton parity automata  with majority memory induces a moderation in  the rate of changing nodes and in 
the damage spreading, albeit in the latter case memory turns out ineffective in the control of the damage
with low memory charge and in networks with high connectivity, i.e., with low $\beta$\,. In a simulation up to $T$ time-steps, 
the maximum depleting effect of memory on the changing rate is achived with a memory length $\tau \simeq$T/2. With very high  memory 
length, the oscillatory-like effect induced by memory in the changing rate is maintained too long to allow a effective depletion
in the changing rate. Thats is so even in low-connected networks, i.e., with large $\beta$\,.

\section*{Acknowledgment}
RAS contribution is partially supported by EPSRC grant EP/H014381/1 and done during a two-months residence in
the University of the West of England, Bristol. AA contribution  is part of the European project 248992
funded under 7th FWP (Seventh Framework Programme) FET Proactive 3: Bio-Chemistry-Based
Information Technology CHEM-IT (ICT-2009.8.3).

\end{document}